# Phase dislocations in hollow core waveguides

**Andrey Pryamikov,***

Prokhorov General Physics Institute of Russian Academy of Sciences; pryamikov@mail.ru
* Correspondence: pryamikov@mail.ru; Tel.: +7 (499) 503 81 93

**Abstract:** This paper discusses the basic concepts of phase dislocations and vortex formation in the electric fields of fundamental air core mode of hollow core waveguides with specific types of rotational symmetry of the core – cladding boundary. Analysis of the behavior of the electric field phase in the transmission bands shows that the mechanism of light localization in the hollow core waveguides with discrete rotational symmetry of the core – cladding boundary cannot be completely described by the ARROW model. For an accurate description of the phase behavior, it is necessary to account for phase jumps of the magnitude of $\pi$ when passing through the phase dislocations.

## 1. Introduction

The problem we focus on in this paper has been under investigation for ten years in the field of hollow core micro – structured optical fibers [1] and it is related to the explanation of the light localization mechanism in the negative curvature hollow core fibers (NCHCFs) [2, 3, 4]. These fibers have been alternatively known in literature as inhibited coupling photonic crystal fibers [5], antiresonant hollow core fibers [6], revolver hollow core fibers [7] and so on. In our opinion, this terminological disparity reflects the fact that the mechanism of light localization is not completely understood. Until now, two main concepts can be developed to explain the strong light localization in these fibers. The first concept is called inhibited coupling mechanism and it assumes a low spatial overlap and strong phase mismatch between transverse phases of the air core modes and the cladding modes of the hollow core waveguides [8]. The second one is based on the well - known ARROW (anti – resonant waveguide) mechanism of light localization [9], which assumes that the walls of the cladding elements of the waveguide can be considered as Fabry - Perrot type resonators. In this case, there are wavelengths associated with large losses in the waveguide when the phase incursion in the wall of the cladding element is equal to $m\pi$ and, accordingly, with small losses when the phase incursion for the fields of the air core mode is equal to $(m+1/2)\pi$, where $m$ is an integer. These hollow core fibers have unique optical properties and various research groups all over the world have achieved outstanding results using these fibers [10, 11].

In this paper, we consider two main reasons for the appearance of phase dislocations and vortices [12] in the fields and energy flows of the fundamental air core mode of hollow core waveguides. In particular, we demonstrate that the vortex of the transverse component of the Poynting vector of the fundamental air core mode centered at the origin is caused by leaky behavior of the fundamental air core mode. In this case, due to singularities of the axial components of electric and magnetic fields of the fundamental air core mode at which $\text{Re}(E_z) = \text{Im}(E_z) = 0$ and $\text{Re}(H_z) = \text{Im}(H_z) = 0$ there is a vortex of energy flow with the center on the axis of the waveguide. A singularity of this type in the energy flows of the fundamental core mode is characteristic of all types of leaky waveguides with the stream lines of the transverse component of the Poynting vector of the fundamental core mode having a spiral shape and, as a result, the azimuthal component of the Poynting vector $P_\varphi \neq 0$. This means that there is an axial component $L_z$ of the kinetic (Abraham - type) total angular momentum density [13] of the leaky fundamental air core mode.

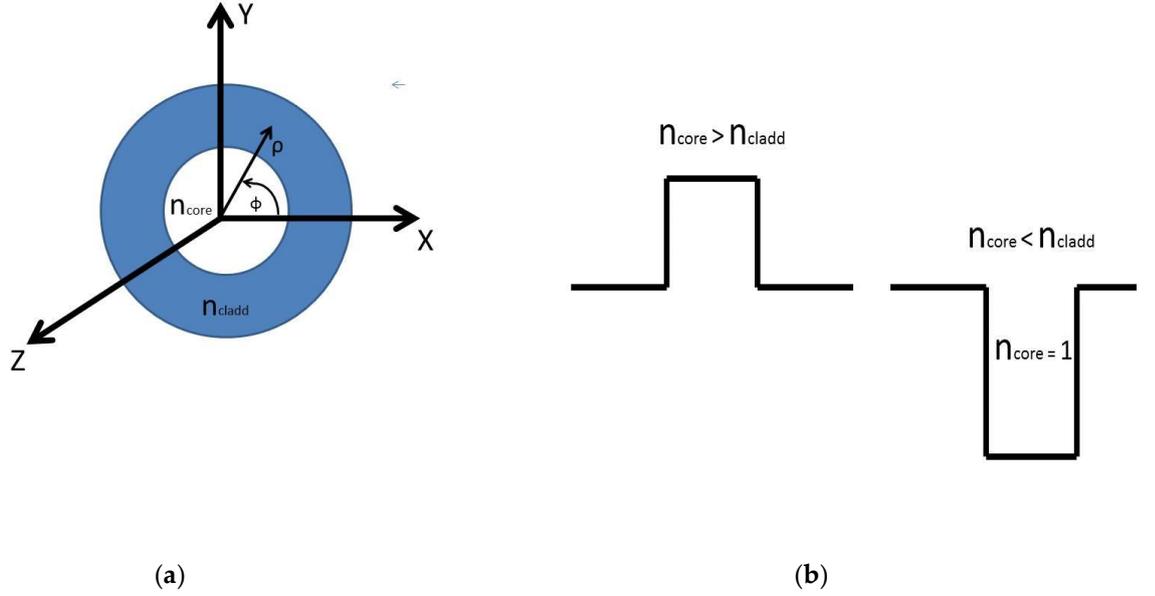

(a)                          (b)

**Figure 1.** (**a**) Schematic representation of the cross –section of a waveguide in a Cartesian and cylindrical coordinate system; (**b**) schematic representation of the refractive index profiles for a waveguide that localizes radiation according to the principle of total internal reflection (step – index fiber) (left) and an air core waveguide with leaky modes (right).

On the other hand, hollow core waveguides, except for hollow core PCFs (photonic crystal fibers) [1], can have a complex shape of the core – cladding boundary in the form of polygons (Kagome lattice hollow core fibers [14]) or have a cladding consisting of a set of capillaries (negative curvature hollow core fibers). In this case, as will be shown below, the formation of singularities and phase dislocations of the electric fields of the fundamental air core mode occur not only at the origin but also in the cladding. If the core – cladding boundary has a discrete rotational symmetry or the elements of the waveguide cladding are arranged according to a specific type of discrete rotational symmetry, the structure of the radial projection of the Poynting vector has the same type of the rotational symmetry. The phase singularities occur in the cladding elements wall along curves obeying the equation $\text{Re}(E_i(x,y)) = \text{Im}(E_i(x,y)) = 0$, where $i$ represents any of the coordinates $(x, y, z)$. When passing through the phase dislocation curve, the phase of the field undergoes a jump of the magnitude of $\pi$, and, therefore, radiation should be reflected from the walls of the cladding elements inside the transmission bands in a different way than according to the ARROW model.

**2. Role of losses in vortex formation in hollow core waveguides**

To demonstrate the physical mechanism of the formation of a central vortex of the transverse component of the Poynting vector located on the z - axis of the hollow core waveguide, we compared the behavior of the transverse component of the Poynting vector of the fundamental core mode for a step – index fiber with no losses and a hollow core waveguide that has an air core in an infinite volume of a dielectric (Fig. 1).

It is known that the axial components of the electric and magnetic fields of hybrid modes of the step – index fiber, localized in the core due to the principle of total internal reflection, can be written as [15]:

$$E_z = A J_n\left(\frac{u}{a}r\right)\cos(n\varphi+\varphi_0),$$

$$H_z = B J_n\left(\frac{u}{a}r\right)\sin(n\varphi+\varphi_0),$$

at $r \leq a$, where $a$ is a radius of the fiber core and:

$$E_z = A \frac{J_n(u)}{K_n(v)} K_n\left(\frac{v}{a}r\right)\cos(n\varphi+\varphi_0),$$

$$H_z = B \frac{J_n(u)}{K_n(v)} K_n\left(\frac{v}{a}r\right)\sin(n\varphi+\varphi_0),$$

at $r > a$. It is assumed that the time dependence and the dependence on the z - coordinate has the form of $e^{i(\omega t - \beta z)}$, $\beta$ is a propagation constant of the core mode. In (1) and (2) $J_n(x)$ and $K_n(x)$ are a Bessel function of the first kind and a modified Bessel function, parameters $u = a\sqrt{k_1^2 - \beta^2}$ and $v = a\sqrt{\beta^2 - k_2^2}$, where $k_1 = \frac{2\pi}{\lambda}n_{core}$, $k_2 = \frac{2\pi}{\lambda}n_{cladd}$ and $\lambda$ is a wavelength, $n_{core}$ and $n_{cladd}$ are refractive indices of the fiber core and cladding. Here, we assume that all refractive indices are real values.

It is known from waveguide electrodynamics that the transverse components of the electric and magnetic fields of the core modes are expressed in terms of axial components as [15]:

$$E_r = -\frac{i}{(k^2 n^2 - \beta^2)}\left(\beta \frac{\partial E_z}{\partial r} + \frac{\omega \mu_0}{r} \frac{\partial H_z}{\partial \varphi}\right),$$

$$E_\varphi = -\frac{i}{(k^2 n^2 - \beta^2)}\left(\frac{\beta}{r} \frac{\partial E_z}{\partial \varphi} - \omega \mu_0 \frac{\partial H_z}{\partial r}\right),$$

$$H_r = -\frac{i}{(k^2 n^2 - \beta^2)}\left(\beta \frac{\partial H_z}{\partial r} - \frac{\omega \varepsilon_0 n^2}{r} \frac{\partial E_z}{\partial \varphi}\right),$$

$$H_\varphi = -\frac{i}{(k^2 n^2 - \beta^2)}\left(\frac{\beta}{r} \frac{\partial H_z}{\partial \varphi} + \omega \varepsilon_0 n^2 \frac{\partial E_z}{\partial r}\right),$$

It can be seen from (1) – (3) that the time averaged Poynting vector $\vec{S} = \frac{1}{2}\text{Re}(\vec{E} \times \vec{H}^*)$ of the fundamental core mode has only one non – zero component. The power carried by the optical fiber moves only along the z – axis and not along $\varphi$ - and $r$ axes in cylindrical coordinate system.

A different behavior can be observed for the simplest hollow core leaky waveguides, which are a round hole and a hollow hexagon in an infinite dielectric layer. Let us consider the vortex properties of the electric and magnetic and the transverse component of the Poynting vector fields of the fundamental air core modes of these waveguides. The numerical calculations were carried out with a COMSOL commercial package at a wavelength of $\lambda$ = 3.39 μm to reduce the number of finite elements in the calculations. The remaining parameters of the hollow core waveguides are shown in the caption of Fig. 2.

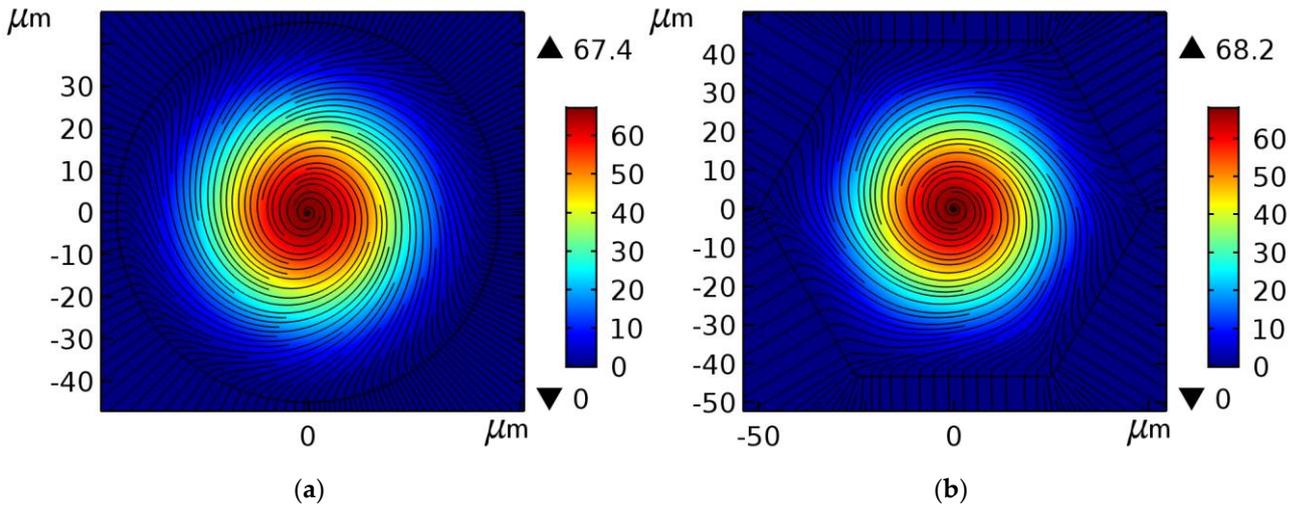

(a)  (b)

**Figure 2.** Axial component of the Poynting vector (in color) and the streamlines of transverse component of the Poynting vector of fundamental air core modes (HE$_{11}$) (black lines) at a wavelength of 3.39 μm: (**a**) air hole in an infinite layer of dielectric (refractive index $n$ = 1.45) with an air core diameter of $D$ = 90 μm; (**b**) air core hexagon in an infinite layer of a dielectric (refractive index $n$ = 1.45) with the same effective mode area of the fundamental air core mode as the one shown in (a).

The behavior of streamlines of the transverse component of the Poynting vector of the fundamental air core mode (Fig. 2) indicates the presence of both the $r$ - and $\varphi$ - projections of the time averaged Poynting vector component and, therefore, the optical vortex . They occur due to different mechanisms of light localization in the air core waveguides and in the step – index fiber (Fig. 1). For the air core waveguide, the presence of waveguide losses leads to occurrence of azimuthal and radial projections of the transverse component of the Poynting vector of the fundamental air core mode. We will consider this behavior using the example of a leaky air core waveguide with a core in the form of a round hole in an infinite layer of a dielectric (Fig. 2a).

The axial components of electric and magnetic fields of a fundamental air core mode (HE$_{11}$ mode) in the air core (Fig. 2a) are described by the same Bessel functions as in the case of a step – index fiber (1). Assuming the fundamental air core mode has a complex propagation constant $\beta = \beta^{(Re)} - i\beta^{(Im)}$, where $\beta^{(Im)} \ll \beta^{(Re)}$ as it often happens in practice, the function argument $u$ in (1) can be represented as $u \approx u^{(Re)} + iu^{(Im)}$, where $u^{(Re)} = a\sqrt{k_1^2 - \beta^{(Re)2}}$ and $u^{(Im)} = a\beta^{(Re)}\beta^{(Im)}$. This expansion of the function argument is possible since, for example, for the fundamental air core mode shown in Fig. 2a

$$\beta = \frac{2\pi}{\lambda}\left(n_{eff}^{(Re)} - in_{eff}^{(Im)}\right) = \frac{2\pi}{\lambda}(0.99958 - i1.5e-5).$$ In this case, it is clear that $\varphi$ - and $r$ - projections of electric and magnetic fields in the air core will have both imaginary and real parts (3). For example, near the origin, the axial components of electric and

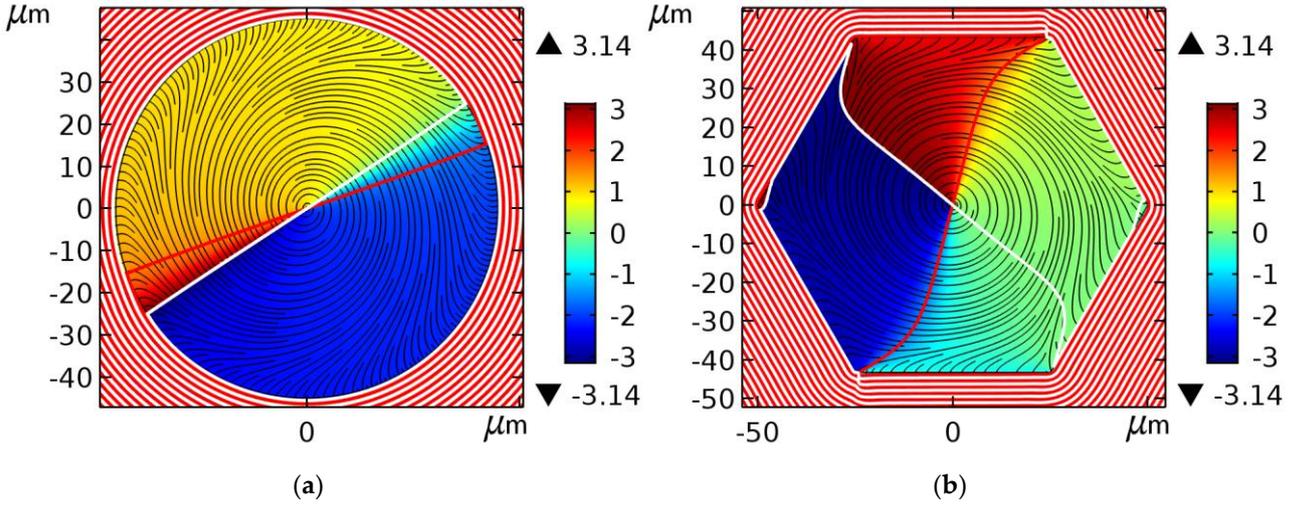

**Figure 3.** Phase distributions of the axial components of electric fields $E_z$ for the air core waveguides shown in Fig. 2: (**a**) the air hole in the infinite layer of a dielectric; (**b**) hollow core hexagon in the infinite layer of a dielectric. Thin red and white lines correspond to the condition that the real (red) and imaginary (white) parts of the axial component of electric field being equal to zero. The phase changes from $-\pi$ (dark blue) to $\pi$ (dark red).

magnetic fields of the fundamental air core mode can be represented using an asymptotic of the Bessel function of $J_1(q)$ at $q \to 0$:

$$E_z \approx 1/2 A q \cos(n\varphi + \varphi_0),$$

$$H_z \approx 1/2 B q \sin(n\varphi + \varphi_0),$$

where $q = (q^{(Re)} + i q^{(Im)}) r$ is a complex argument and $q^{(Re)} = \sqrt{k_1^2 - \beta^{(Re)2}}$, and $q^{(Im)} = \beta^{(Re)} \beta^{(Im)}$. The axial components of electric and magnetic fields of the fundamental air core mode have the same structure (4) and they are proportional to $q$ (4). This means that the real and imaginary parts of the axial components of electric and magnetic fields simultaneously tend to zero and $\text{Re}(E_z) = \text{Im}(E_z) = \text{Re}(H_z) = \text{Im}(H_z) = 0$ at $r = 0$ forming a vortex (Fig.3a). The transverse component of the time averaged Poynting vector of the fundamental air core mode will have non – zero $\varphi$ - and $r$ - projections

$$\vec{S}_{transv} = \frac{1}{2} \text{Re}\left[ \left( E_\varphi H_z^* - E_z H_\varphi^* \right) \vec{r} + \left( E_z H_r^* - E_r H_z^* \right) \vec{\varphi} \right] \neq 0$$

at any point of the waveguide cross – section, except for the origin (Fig. 2a). At the origin, both projections of the transverse component of the Poynting vector component are equal to zero. Thus, the leaky fundamental air core mode has the kinetic (Abraham - type) angular momentum density [13].

For the hollow core hexagon (Fig.2b), the axial components of electric and magnetic fields of the fundamental air core mode can be represented as series of Bessel functions of the first kind of different orders with a corresponding asymptotic at the origin as in (4). The propagation constant of the fundamental air core mode is $\beta = \frac{2\pi}{\lambda}(0.99958 - i1.53e-5)$ and the vortex formation is confirmed by the phase distribution of the axial electric fields of the mode (Fig. 3b). It means that, as in the case of an air hole waveguide, real and imaginary parts of the axial components of the electric and magnetic fields are equal to zero at the origin and there is a vortex of the transverse component of the Poynting vector of the fundamental air core mode (Fig.2b). Both phase distributions (Fig. 3) are governed by a nonlinear rule [16] and these vortices are called anisotropic. In our work [17], we showed that the coordinates of the vortex centers of the transverse component of the Poynting vector of the leaky core modes also satisfy the equation $P_x(x,y) = P_y(x,y) = 0$, where $P_x$ and $P_y$ are projections of the transverse component of the Poynting vector.

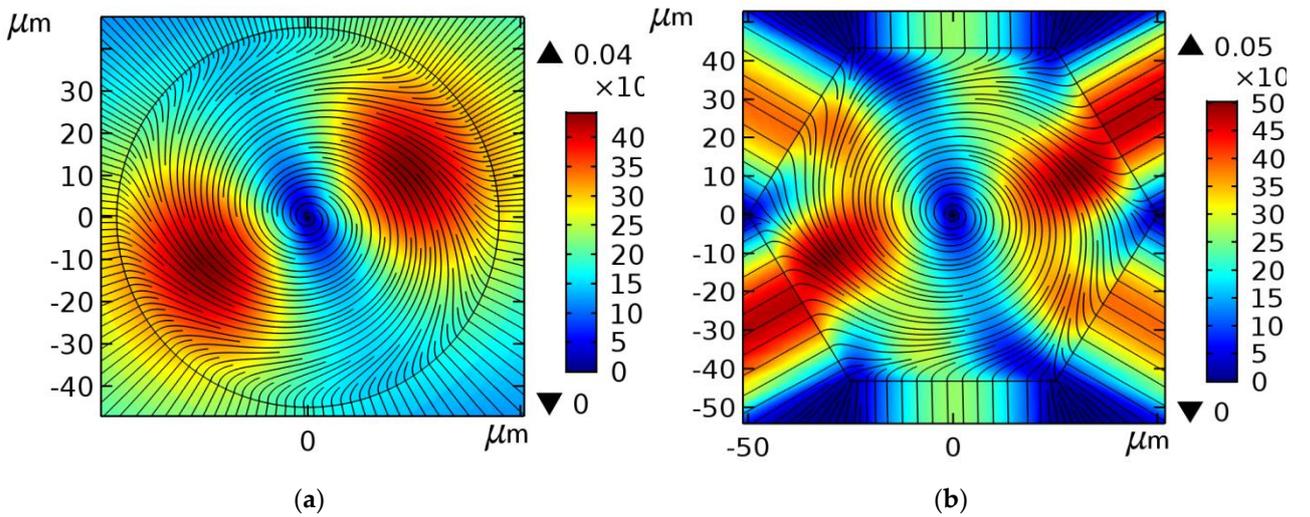

**Figure 4.** Radial projection of the transverse component of the Poynting vector of the fundamental air core mode at a wavelength of 3.39 μm for two air core waveguides: (**a**) air hole in an infinite dielectric layer (Fig. 2a); (**b**) air core hexagon in an infinite dielectric layer (Fig. 2b).

**3. Impact of the core – cladding boundary shape on the vortex formation. Problem of loss reduction**

In this section, we would like to consider the correlation between the vortex formation in the transverse component of the Poynting vector and phase dislocations of the fields of the fundamental air core mode, not only at the origin, but also in the waveguide cladding. Here it is important to determine of light localization mechanism in hollow core waveguides with a complex shape of the core – cladding boundary and to reduce respective waveguide losses.

To better understand the difference between the processes of the energy leakage in hollow core waveguides with various rotational symmetries of the core – cladding boundary, we consider two examples. In the first example, we examine the distribution of the radial projection of the transverse component of the Poynting vector of the fundamental air core mode in two waveguides discussed in the previous section. The results of the numerical calculation are shown in Fig. 4. In both cases shown in Fig. 4, the sign of the radial projection of the transverse component of the Poynting vector changes from zero to some positive values, taking no negative values. This is due to the fact that in both cases, the transverse component of the Poynting vector has only one vortex located at the origin. In addition, the radial projection of the transverse component of the Poynting vector in the case of a hollow core hexagon has an inhomogeneous distribution in the waveguide cross – section not related to the discrete rotational symmetry of the core – cladding boundary. The losses in both cases are approximately equal as indicated in the previous section (the imaginary parts of the propagation constants).

However, similar waveguides with a finite wall thickness show a completely different behavior. Let the waveguides considered above be located in empty space and have a wall thickness of 1 μm, so that the radiation is transmitted in the longest wavelength transmission band. For the capillary the distribution of the radial projection of the transverse component of the Poyning vector of the fundamental air core mode does not change qualitatively compared to the case shown in Fig.4a (Fig. 5a). Only the imaginary part of the propagation constant decreases approximately by two orders of magnitude $\beta = \frac{2\pi}{\lambda}(0.99958 - i7.7e-7)$.

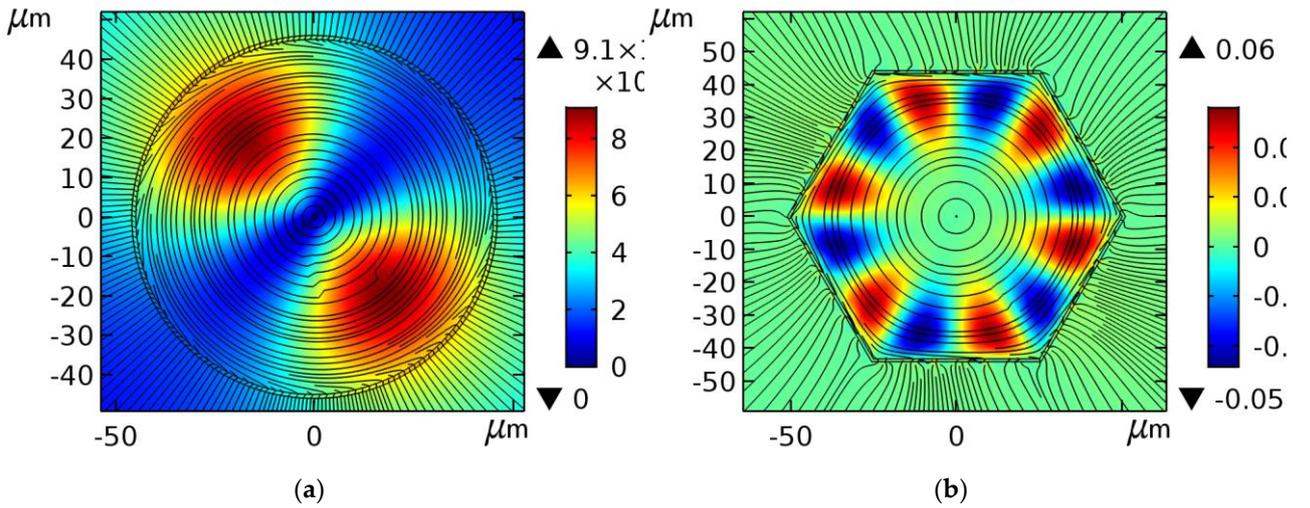

**Figure 5.** Radial projection of the transverse component of the Poynting vector of the fundamental air core mode at a wavelength of 3.39 μm for: (**a**) a capillary with a wall thickness of 1 μm (refractive index $n$ = 1.45) with an air core diameter of $D$ = 90 μm ; (**b**) hollow hexagon with a wall thickness of 1 μm (refractive index $n$ = 1.45). Black lines are streamlines of the transverse component of the Poynting vector of the fundamental air core mode. The effective mode areas are the same in both cases.

For the hollow core hexagon, the structure of the radial projection of the transverse component of the Poyning vector of the fundamental air core mode has, in contrast to the distribution shown in Fig. 4b, a certain discrete rotational symmetry that coincides with the symmetry of the core – cladding boundary. It should be noted that, as it was shown in our work [18], in contrast to the capillary (Fig. 5a) the radial projection of the transverse Poynting vector component for the hollow core polygon has both positive (dark red) and negative (dark blue) values (Fig. 5b). For the hollow core hexagon, the imaginary part of the propagation constant is somewhat lower than for the capillary $\beta = \frac{2\pi}{\lambda}(0.99958 - i5.4e-7)$.

To qualitatively explain why the hollow hexagon with a finite wall thickness has the structure of the radial projection of the transverse component of the Poynting vector shown in Fig. 5b let us consider refraction and reflection of radiation from the core – cladding boundary. For the waveguides shown in Fig. 4 there is a quasi – standing wave (the air core mode) and refracted waves at the core boundary running away to infinity outside the waveguides. It is known that for an optical vortex to form under the interference of plane waves, it is necessary to have at least three waves in the considered space region [19]. Therefore, vortices in electromagnetic fields and energy flows of the fundamental air core mode for the waveguides shown in Fig.4 can be formed only at the origin, as it was shown in Section 2. A similar reasoning can be applied to the case of the capillary (Fig. 5a), namely, in addition to the air core mode and the wave going to infinity outside the waveguide, there are only two cylindrical waves in the capillary wall. For the axial components of electric and magnetic fields of the fundamental air core mode, they can be expressed as the sum of two Hankel functions:

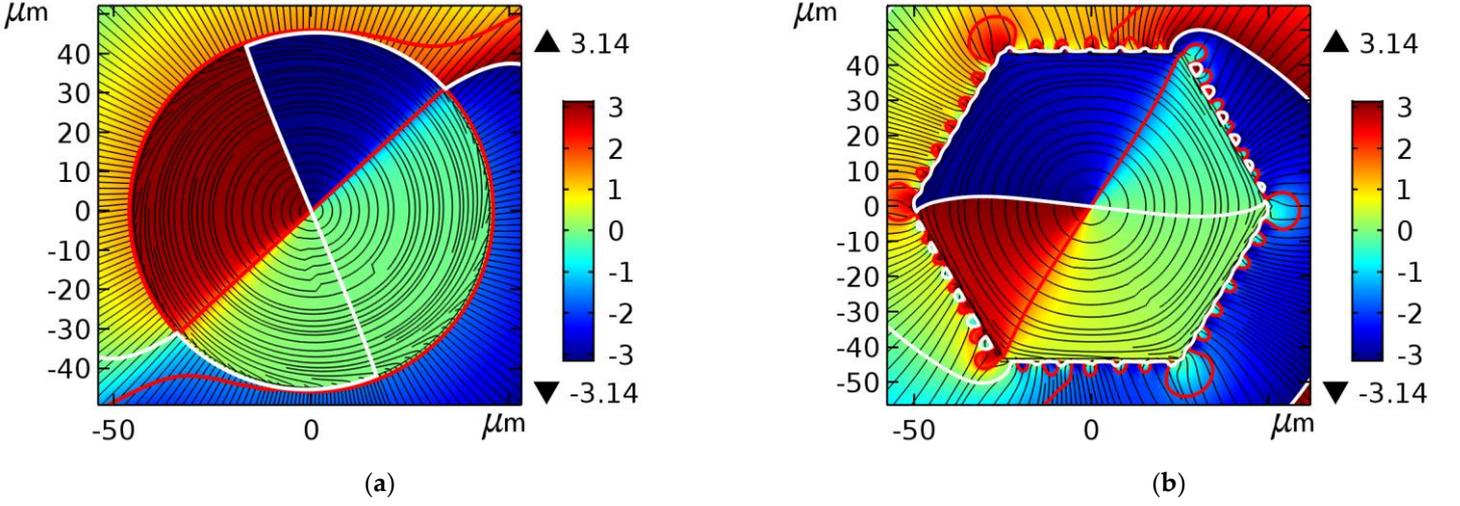

(a)          (b)

**Figure 6.** Phase distribution of the axial electric field component at a wavelength of 3.39 μm for: **a)** a capillary with a wall thickness of 1 μm (**b**) hollow hexagon with a wall thickness of 1 μm. Black lines are streamlines of the transverse component of the Poynting vector of the fundamental air core mode. Thin red and white lines correspond to the condition that the real (red) and imaginary (white) parts of the axial component of electric field being equal to zero. The phase changes from −π (dark blue) to π (dark red).

$$E_z = \left(AH_1^{(1)}(k_t r) + BH_1^{(2)}(k_t r)\right)\cos(\varphi+\varphi_0),$$

$$H_z = \left(CH_1^{(1)}(k_t r) + DH_1^{(2)}(k_t r)\right)\sin(\varphi+\varphi_0),$$

where $k_t = \dfrac{2\pi}{\lambda}\sqrt{n_{wall}^2 - \beta^2}$, where $n_{wall}^2$ is a refractive index of the capillary wall. The optical vortices in the axial component of electric fields occur at $\text{Re}(E_z) = \text{Im}(E_z) = 0$ and in the energy flows when $P_x = P_y = 0$ ($P_x$ and $P_y$ are projections of the transverse Poynting vector component) and don't occur in the core – cladding capillary wall and, therefore, their formation is possible only at the origin.

For the hollow core hexagon, the presence of a finite thickness of the core – cladding boundary combined with angular regions of the wall leads to a complex interference inside the waveguide wall, in which the axial components of electric and magnetic fields of the fundamental air core mode are described by series:

$$E_z = \sum_n \left(A_n H_n^{(1)} + B_n H_n^{(2)}\right)\cos(n\varphi+\varphi_0),$$

$$H_z = \sum_n \left(C_n H_n^{(1)} + D_n H_n^{(2)}\right)\sin(n\varphi+\varphi_0),$$

where $A_n$, $B_n$, $C_n$ and $D_n$ are the harmonic amplitudes.

The phase distributions of the axial component of the electric field of the fundamental air core mode for the capillary and hollow core hexagon are shown in Fig. 6. The Fig. 6a shows that the distribution of the axial component of the electric field of the fundamental air core mode has only one phase singularity located on the axis of the waveguide. The difference between the phases of the axial component of electric field in the hollow core and in the outer space is approximately 1.6 radians at each point of the capillary wall. This value is consistent with the values of the phase difference obtained from the ARROW model, when the phase difference should be equal to π/2 for the antiresonant condition in the longest wavelength transmission band. Also, there are no phase inhomogeneities along the perimeter of the capillary. For the hollow core hexagon (Fig. 6b), the phase distribution of the axial component of the electric field of the fundamental air core mode is inhomogeneous along the perimeter of the hexagon. The phase distribution is periodic in accordance with the discrete rotational symmetry of the core – cladding boundary and the phase difference between the field inside the air core and in the outer space varies in adjacent sections of the waveguide wall from exact value of π to arbitrary values of radians lying in the range from 0 to π/2 radians. It can be shown that the phase of transverse components of the electric field of the fundamental air core mode behaves in the same way. At the same time, the losses for the hollow hexagon, as it follows from the values of the imaginary parts of the propagation constants of the fundamental air core mode, are slightly lower than for the capillary.

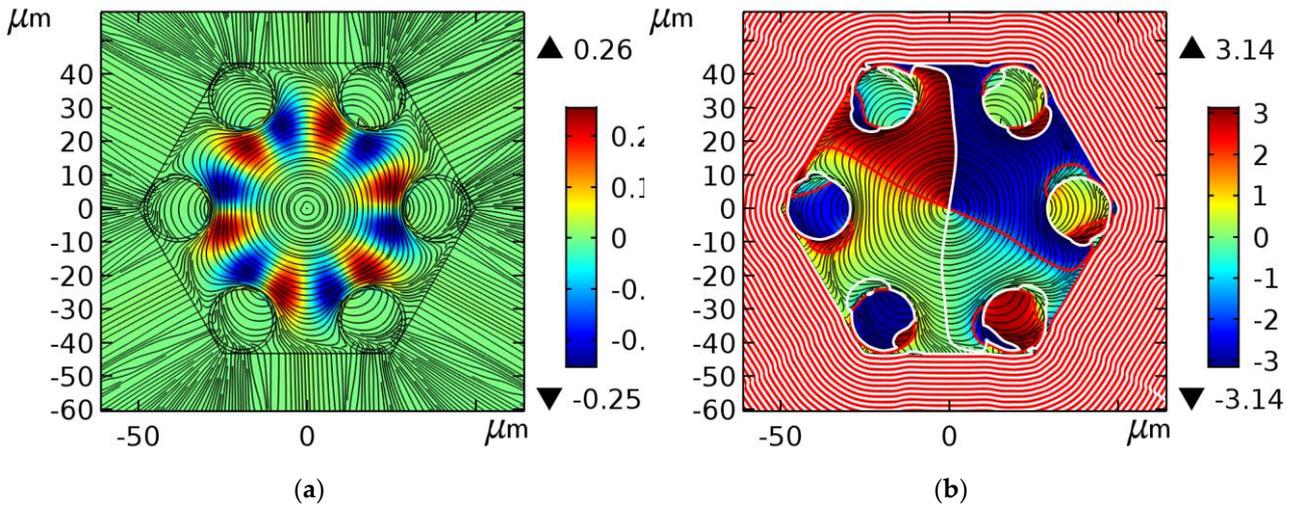

**Figure 7.** Hollow core waveguides formed by six capillaries with an outer diameter of 20 μm and a wall thickness of 1 μm inserted into the corners of the hollow core hexagon (Fig. 2b): (**a**) radial projection of the transverse component of the Poynting vector of the fundamental air core mode at a wavelength of 3.39 μm; (**b**) phase distribution of the axial electric field component of the fundamental air core mode at a wavelength of 3.39 μm. Thin red and white lines correspond to the condition that the real (red) and imaginary (white) parts of the axial component of the electric field being equal to zero. The phase changes from −π (dark blue) to π (dark red). The refractive index of the capillaries is 1.45.

It can be seen from the above that this a difference in the interaction of the fundamental air core mode with the core – cladding boundary wall for the capillary and the air core hexagon should lead to different types of light reflection from the waveguide wall and the air core mode energy leakage. For the capillary, the effective light localization occurs under the antiresonant condition $2\pi d/\lambda \sqrt{n_{wall}^2 - \beta^2} = (m+1/2)\pi$, where $d$ is the thickness of the core – cladding boundary wall and $m$ is an integer corresponding to the number of transmission bands (in our case $m=0$ and the phase shift is π/2). This phase shift occurs along the entire core – cladding boundary of the capillary. It is well known from singular optics that when passing through the phase dislocation of electromagnetic fields, the phase abruptly changes by π [20]. For the hollow core hexagon, the phase jump of π for the electric field of the air core fundamental mode inside the air core and outside the waveguide occurs due to phase dislocations arising in the core – cladding boundary wall at the core – cladding boundary sections where $\mathrm{Re}(E_z(x,y)) = \mathrm{Im}(E_z(x,y)) = 0$ and $\text{Re}\left(E_i^{transv}(x,y)\right) = \text{Im}\left(E_i^{transv}(x,y)\right) = 0$, where $i = \{x, y\}$. The light reflection from the waveguide walls occurs precisely in this phase regime and is not consistent with the ARROW model.

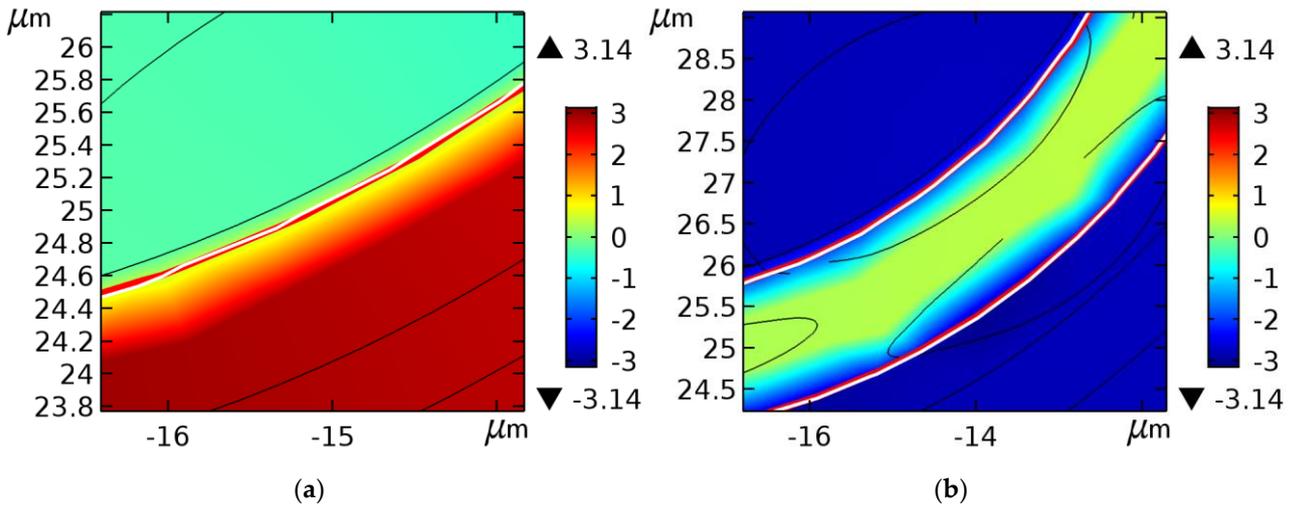

**Figure 8.** Phase distribution of the axial component of the electric field of the fundamental air core mode in the cladding capillary wall at a wall thickness: (**a**) $d = 1$ μm (the longest wavelength transmission band); (**b**) $d = 2.42$ μm. The phase changes from $-\pi$ (dark blue) to $\pi$ (dark red), with light green corresponding to zero. Thin red and white lines correspond to the condition that the real (red) and imaginary (white) parts of the axial component of electric field being equal to zero. For (**a**) red and white lines superimpose on the inner border of the cladding capillary. Other parameters are the same as in Fig. 7.

To confirm this finding, let us consider a well – known hollow core waveguide [21] with a cladding consisting of six capillaries of the same thickness as the wall thickness of the waveguides discussed above (Fig. 5). The capillaries will be located at the vertices of the hollow core hexagon shown in Fig. 2b. Figure 7a shows the distribution of the radial projection of the transverse component of the Poynting vector of the fundamental air core mode of the waveguide. As can be seen, the distribution exactly repeats the similar distribution obtained for the hollow core hexagon (Fig. 5b) and does not coincide with the distribution in Fig. 4b. The radial projection of the transverse component of the Poynitng vector has both positive and negative values as in Fig. 5b. This suggests that the presence of thin – walled capillaries at the vertices of the hexagon leads to the same light localization mechanism as for the hollow core hexagon (Fig. 5b). In addition, the effective mode area of the waveguide shown in Fig. 7 is smaller than in the case shown in Fig. 5b, but the propagation constant of the fundamental air core mode is $\beta = 0.99916 - i6.6e - 7$. The loss values are very close to each other.

To confirm that the radiation is localized in the hollow core waveguide (Fig. 7a) due to the presence of phase dislocations of the fields of the fundamental air core mode, the phase distribution of the axial component of electric field $E_z$ was plotted in Fig. 7b. It can be seen from Fig. 7b that the phase distribution of the axial component of the electric field has a vortex structure and a singularity at the origin. Also, the phase experiences a jump equal to $\pi$ between the field inside the cladding capillary and outside it. This suggests that in this case the mechanism of radiation reflection and its localization in the air core is the same as for the hollow core hexagon (Fig. 6b). The difference is that for the waveguide with a cladding consisting of capillaries the phase jump equal to $\pi$ occurs only in the capillary wall. Figure 8a shows enlarged image of the phase distribution of the axial component of the electric field (Fig. 7b) near and inside the capillary wall in the longest wavelength transmission band. Figure 8b shows the same distribution for the cladding capillary wall thickness of 2.42 μm (according to the ARROW model, the wavelength of 3.39 μm is antiresonant for $m=1$ at this value of the capillary wall thickness). The figures clearly show the phase jumps of the magnitude of π when passing through the phase dislocations of the electric field.

**5. Discussion**

In this work, the mechanisms of light localization of the fundamental air core mode in hollow core waveguides were considered from the perspective of the phase dislocations of the mode fields and their impact on vortex formation in the transverse component of the Poynting vector. According to the ARROW model, the most efficient reflection of radiation of the air core mode from the waveguide wall can be achieved at an antiresonant condition when phase incursion in the waveguide wall is $(m+1/2)\pi$, where $m$ is an integer. In this case, the number of lines of zero values of any component of the electric field of the fundamental air core mode in the core – cladding boundary wall determines the number of the transmission band of the waveguide. For example, the phase incursion in the capillary wall in the second transmission band is $3\pi/2$ and two lines of zero values of the axial or transverse component of electric field are in the capillary wall according to the ARROW model. In this case, the imaginary or real part of the electric field components is not specified. While this model accurately describes the light localization in waveguides with a continuous rotational symmetry of the core – cladding boundary, it should always be borne in mind that the components of the electric fields of the air core mode have both imaginary $\text{Im}(E(x,y))=0$ and real part $\text{Re}(E(x,y))=0$ that have their own lines of zero values in space. For a single capillary, lines of zero values of axial component of electric field intersect with each other only at the origin. A completely different behavior occurs in hollow core waveguides with a finite wall thickness comparable to the wavelength and with a specific discrete rotational symmetry of the core – cladding boundary. In this case, the lines of zero values of the real and imaginary parts of the axial component of the electric field can intersect and overlap not only at the origin but also in the cladding forming phase dislocations of different types. The phase changes abruptly to π when passing through this phase dislocation of the air core mode electric fields and its behavior cannot be explained by the ARROW model. This behavior of the fundamental air core mode fields in the cladding can generate the corresponding vortices in the transverse component of the Poynting vector [17]. In our opinion, all of the above consideration should be taken into account for a correct understanding of light localization mechanism in hollow core waveguides and application of the ARROW model.

**Funding:** This research was funded by Russian Science Foundation, grant number 19 – 12 - 00134.

**Data Availability Statement:** The data presented in this study are available on request from the corresponding author. The data are not publicly available due to the fact that the most of the data was obtained using a commercial package COMSOL.

**Conflicts of Interest:** The authors declare no conflict of interest.

**References**
1. Russell, Philip. St. J. Photonic – crystal fibers. *J. Light. Technol.* **2006**, *24*, 4729 – 4749.

2. Yu, Fei and Knight Jonathan C. Negative curvature hollow – core optical fibers. *IEEE J. Sel. Top. Quantum Electron*. **2016**, *22*, 4400610.
3. Wei, C; Weiblen, R. Joseph; Menyuk, Curtis. R.; and Hu, J. Negative curvature fibers. *Adv. Opt. Photonics* **2017**, *9*, 504 – 561.
4. Pryamikov, A. D.; Biriukov, A. S.; Kosolapov, A. F.; Plotnichenko, V. G.; Semjonov, S. L.; and Dianov, E. V. Demonstration of a waveguide regime for a silica hollow – core microstructured optical fiber with a negative curvature of the core boundary in the spectral region > 3.5 μm, *Opt. Express* **2011**, *19*, 1441 – 1448.
5. Debord, B.; Alharbi, M.; Bradley, T.; Fourcad – Dutin, C; Wang, Y. Y.; Vincetti, L.; Gérôme, F.; and Benabid. F. Hypocycloid – shaped hollow – core photonic crystal fiber Part I: Arc curvature effect on confinement loss, *Opt. Express* **2013**, *21*, 28597 – 28608.
6. Poletti, F. Nested antiresonant nodeless hollow core fiber, *Opt. Express* **2014**, *22*, 23807 – 23828.
7. Bufetov, I. A.; Kosolapov, A. F.; Pryamikov, Gladyshev, A. V.; Kolyadin, A. N.; Krylov, A. A.; Yatsenko, Y. P.; Biriukov, A. S. Revolver hollow creo optical fibers, *Fibers* **2018**, 6, 39 – 65.
8. Debord, B.; Amsanpally, A..; Chafer, M.; Baz, A.; Maurel, J.; Blongy, J. M.; Hugonnot, E.; Scol, F.; Vincetti, L.; Gérôme, F.; and Benabid. F. Ultralow transmission loss in inhibited – coupling guiding hollow fibers, *Optica* **2017**, *4*, 209 - 217.
9. White, T. P.; McPhedran, R. C.; Sterke, C. M.; Litchinitser, N. M.; Eggleton, B. J. Resonances in microstructured optical waveguides, *Opt. Express*, **2003**, *11*, 1243 – 1251.
10. Chafer, M.; Osôrio, J. H.; Amrani, F.; Delahaye, F.; Maurel, M.; Debord, B.; Gérôme, F.; and Benabid. F. 1 – km Hollow – core fiber with loss at the silica Rayleigh limit in the green spectral range, *IEEE Photonics Technol. Lett.* **2019**, *31*, 685 – 688.
11. Jasion, G. T.; Bradley, T.; Sakr, H.; Hayes, J. R.; Chen, Y.; Taranta, A.; Mulvad, H. C.; Davidson, I. A.; Wheeler, N. V.; Fokoua, E. N.; Wang, W.; Richardson, D. J.; Poletti, F. Recent breakthrough in hollow core fiber technology, Proceedings of SPIE OPTO, San Francisco, California, United States, **2020**, 1130902.
12. Gbur, G. J. *Singular Optics*, CRC Press, Taylor&Francis Group, **2017**, 33 – 60.
13. Picardi, M. F.; Bliokh, K. Y.; Rodriguez – Fortun, F. J.; Alpeggiani, F.; Nori, F. Angular momenta, helicity, and other properties of dielectric – fiber and metallic – wire modes, *Optica*, **2018**, *5*, 1016 – 1024.
14. Couny, F.; Benabid, F.; Roberts, P.J.; Light, P.S.; Raymer, M. G. Generation and photonic guidance of multi – octave optical – frequency combs, *Science* **2007**, 1118 – 1121.
15. Katsunari Okamoto. *Fundamentals of optical waveguides*, Elsevier Inc., 2006, 58 – 70.
16. Mokhun, I. I. Introduction to linear singular optics. In *Optical Correlation, Technique and Applications,* Editor Oleg V. Angelsky; SPIE Press: Bellingham, Washington USA, 2007; 1 – 133.
17. Pryamikov, A.; Alagashev, G.; Falkovich, G.; Turitsyn, S. Light transport and vortex – supported wave – guiding in micro – structured optical fibres, *Sci. Rep*. **2020**, 10: 2507.
18. Pryamikov, A. D.; Alagashev, G. K. Strong light localization and a peculiar feature of light leakage in the negative curvature hollow core fibers, *Fibers* **2017**, *5*, 43.
19. Rozanov, N. N. Formation of radiation with wave – front dislocations. *Opt. Spectrosc*. **1993**, *75*, 861 – 867.
20. Soskin, M. S.; Vasnetsov, M. V. Singular Optics. In *Progress in Optics*, Editor Emil Wolf, Elsevier, 2001, Volume 42, 219 – 276.
21. Uebel, P.; Günendi, M.C.; Frosz, M. H.; Ahmed, G.; Edavalath, N. N.; Ménard, Jean – Mishel; Russell, Philip St. J. Broadband robustly single – mode hollow core PCF by resonant filtering of higher – order modes. *Opt. Lett* **2016**, *41*, 1961 – 1964.